\documentclass{article}
\usepackage[english]{babel}
\usepackage{graphicx}
\usepackage{caption}
\usepackage{multirow}
\usepackage{rotating}
\newcommand{\be}{\begin{equation}}
\newcommand{\ee}{\end{equation}}
\newcommand{\bea}{\begin{eqnarray}}
\newcommand{\eea}{\end{eqnarray}}

\begin{document}
 \title{\bf Determination the Parameters of Markowitz Portfolio Optimization Model}
\author {
              {\bf  Ertugrul BAYRAKTAR
}\\
        {\it  Graduate School of Science and Engineering, Mechatronics Engineering}\\
        {\it Istanbul Technical University}\\
        {\it Mekatronik Egitim ve Arastirma Merkezi }\\
       {\it Istanbul Teknik Universitesi, Maslak Kampusu $34469$ Maslak}\\
   {\it  Istanbul, Turkey}\\
        {\it  e-mail:  bayraktare@itu.edu.tr}\\
        {\it  Tel: $0090(212) 285 7005$ Fax: $0090(212) 328 0337$ }\\{} \\
        {\bf  Ay\c{s}e H\"{u}meyra B\.ILGE}\\
        {\it  Faculty of Engineering and Natural Sciences, Kadir Has University}\\
        {\it  Kadir Has Universitesi, Cibali, $34083$, Fatih}\\
        {\it  Istanbul, Turkey}\\
        {\it  e-mail: ayse.bilge@khas.edu.tr}\\
        {\it  Tel: $0090(212) 532 6532$ Fax: $0090(212) 533 2286$ }\\{} \\
          }
\maketitle
\begin{abstract}
\baselineskip=10pt \noindent The main purpose of this study is the
determination of the optimal length of the historical data for the
estimation of  statistical parameters in Markowitz Portfolio
Optimization. We present a trading simulation using Markowitz
method, for a portfolio consisting of foreign currency exchange
rates and  selected assets from  the Istanbul Stock Exchange ISE
30, over the period $2001$-$2009$. In the simulation, the expected
returns and the covariance matrix are computed from historical
data observed for past $n$ days and the target returns are chosen
as multiples of the return of the market index.  The trading
strategy is to buy a stock if the simulation resulted in a
feasible solution and sell the stock after exactly $m$ days,
independently from the market conditions. The actual returns are
computed for $n$ and $m$  being equal to $21$, $42$, $63$, $84$
and $105$ days and we have seen that the best return is obtained
when the observation period is $2$ or $3$ times the investment
period.
\end{abstract}

\smallskip
\baselineskip=10pt \noindent \textbf{Keywords:}  Markowitz Portfolio Optimization, Observation and Investment Period, Sliding Window Application to ISE 30, Parameter Determination.

\baselineskip=15pt
\section{ Introduction}
Portfolio optimization is the process of investing to financial
instruments with the aim of optimizing certain criteria subject to a number of equality or inequality constraints.  The
output of the optimization process is a distribution of weights of these instruments in the portfolio. Markowitz portfolio
optimization theory \cite{1} is considered to be the
milestone of modern finance theory. In the basic application of
the model, the aim is to achieve minimum risk subject to a
specified level of expected return (or  maximum return subject to a specified level of calculated risk) \cite{2}.

The mathematics underlying the Markowitz portfolio optimization is
the standard quadratic optimization problem subject to equality and
inequality constraints. This is a well known problem that has
implementations on many platforms. In the minimization of the
risk for a predetermined level of target return, the objective
function is the ``risk" which is assumed to be represented by the quadratic form
$\frac{1}{2} X^t Q X$
where $Q$ is the covariance matrix of the assets in the portfolio and $X$ is the solution vector
that will give the weights of these assets in the portfolio.  The crucial constraint expresses
the requirement that the portfolio return $R^tX$ be at least equal to a target return.
$R^tX\ge R_0.$
There may be other constraints that reflect the investors
preferences or obligations towards certain groups of assets. Thus,
once the covariance  matrix $Q$ and the expected return vector $R$
have been chosen, the problem is straightforward. However, the
optimal solution is very sensitive to these parameters and the
determination of the covariance matrix and the expected return
vector is crucial. Thus although the basic method is old and well
established, the application of the method is still a current
research interest. In the literature, this problem is addressed
either by a reformulation of the problem \cite{3},
by including  the estimation risk into the problem \cite{4},
or by including the indeterminacy of the parameters into
the model and using conic programming \cite{5} and also by including
the estimation of optimal portfolio return \cite{6}. We also note that
the portfolio optimization problem can be solved as a linear optimization instead of quadratic optimization \cite{7} .

In earlier investigations we noticed that the statistical parameters
of market data were very sensitive to the
length of the observation period and we decided to conduct a study
for a quantitative measure of the effect of these fluctuations to overall
performance of the method.  We used historical data of various lengths to estimate
the covariance matrix and the expected return vector;  with the purpose of determining
the optimal relative lengths of the past data to be
used for parameter estimation  and the mean trading period.

For this purpose, we applied the Markowitz Porfolio Optimization method  to selected assets
from the Istanbul Stock Exchange ISE 30, over the period $2001$-$2009$ and
we chose an investment strategy displaying the effects of the relative lengths of
the observation and investment periods.
  This investment strategy is as follows: {\it ``Each day, if the ISE 30 index is down, no purchase is made.
  If it is up, a  buy operation, targeting $k=2$, $3$ and $10$ times  market average, is made. This purchase is based on the output
  of the Markowitz Porfolio Optimization, whose parameters are obtained from historical data for past $n$ days.
  The portfolio is sold after exactly $m$ days, regardless of the market conditions and the net return is computed.
  This process is repeated each day."}

We applied this strategy to the data covering the period 2001-2009
for a   range of observation and investment periods.  The data and methodology is described
in Section 2.  The dependency of the
parameters on the observation length is discussed in Section 3.
 The investment strategy and simulation results are presented in Section 4.
\section{Data and Methodology}
The Markowitz portfolio optimization method is a
constrained optimization problem where the objective function is
the risk of the portfolio \cite{8}; in our investment strategy the
inequality constraints force the portfolio return to be higher
than a targeted return and the equality constraints consist of
various technical restrictions.  This is a typical quadratic
constrained optimization problem that can be solved with
standard computational tools. We have used $14$ stocks
from ISE 30, and the exchange rates for Euro and USD, a total of
$16$ assets.  Denoting the daily closing prices by $P_i$, the
daily returns $R_i$ and the daily logarithmic returns $r_i$  are
defined respectively as below.
$$R_i=\frac{P_i-P_{i-1}}{P_i},\quad\quad r_i=\ln (1+R_i).$$
In the following we work with the logarithmic return series for all assets.
\subsection{Overview of the Assets}
We start by a general classification of the assets under
consideration based on their expected return versus risk. For
this, we compute the mean and the standard deviation of the
logarithmic returns over the whole period as presented in Figure
1. From the figure, we can clearly see that the assets form
clusters that we have denoted as groups $1$-$5$.
\begin{figure}[h]
\begin{center}
\includegraphics[width=0.65\textwidth]{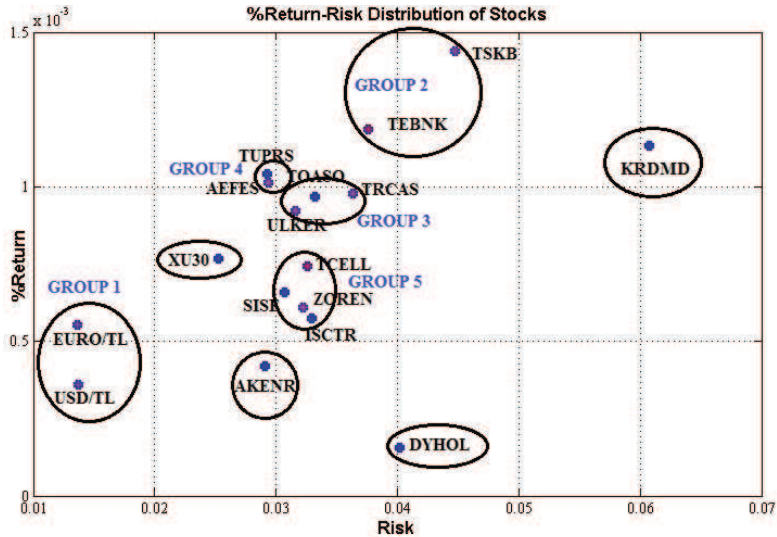}
\caption{Return-Risk distribution of investment instruments.  The first group includes currencies and
it has low risk and low return values. Group 2 includes stocks from banking industry and it has the highest risk and
highest return values excluding KRDMD which seem like an outlier. Group 3 and Group 4 consist of industrial companies and they have higher
return but lower/same risk compared with Group 5 which consists Telecommunication, Manufacturing, Energy companies and a Bank.}
\end{center}
\end{figure}

As expected, the currencies (USD and Euro) form the ``low risk-low return" group, denoted as Group 1.
In this group,  EURO and USD have the same risk value but EURO has higher return value than USD.
The next group is the highest return cluster, Group 2, consisting of the banking stocks TEBNK and TSKB.
This group has higher risk and higer return compared to the ISE 30 index.  The risk and the return of the
stocks which are in the groups of $3$, $4$ and $5$ are closer to the  values of the ISE 30 index.  Except
for ISCTR and TCELL, these groups consist of  industrial manufacturing companies or energy corporations.
The ISE 30 market index belongs to Group 4.
The  KRDMD and DYHOL stocks that have either exceptionally high return-high risk or low return-moderate
risk do not belong to any of these groups.

\section{Statistical Analysis}

In this section we shall study the statistical properties of the data. It is well known that the raw stock prices are non-stationary
but logarithmic returns over relatively short periods of time are usually considered as stationary.  We display in Figure 2,
the logarithmic returns for stocks, currencies, the market index and the raw returns for all assets.  The raw returns have
clearly a rising trend over the second half of the decade and display a depression during the period of 2008 crisis.
On the other hand, due to scaling,  the effects of 2001 crisis are not that obvious in the price data while the are clearly
observed in the return data.
\begin{figure}[h]
\begin{center}
\includegraphics[width=400pt,keepaspectratio=true]{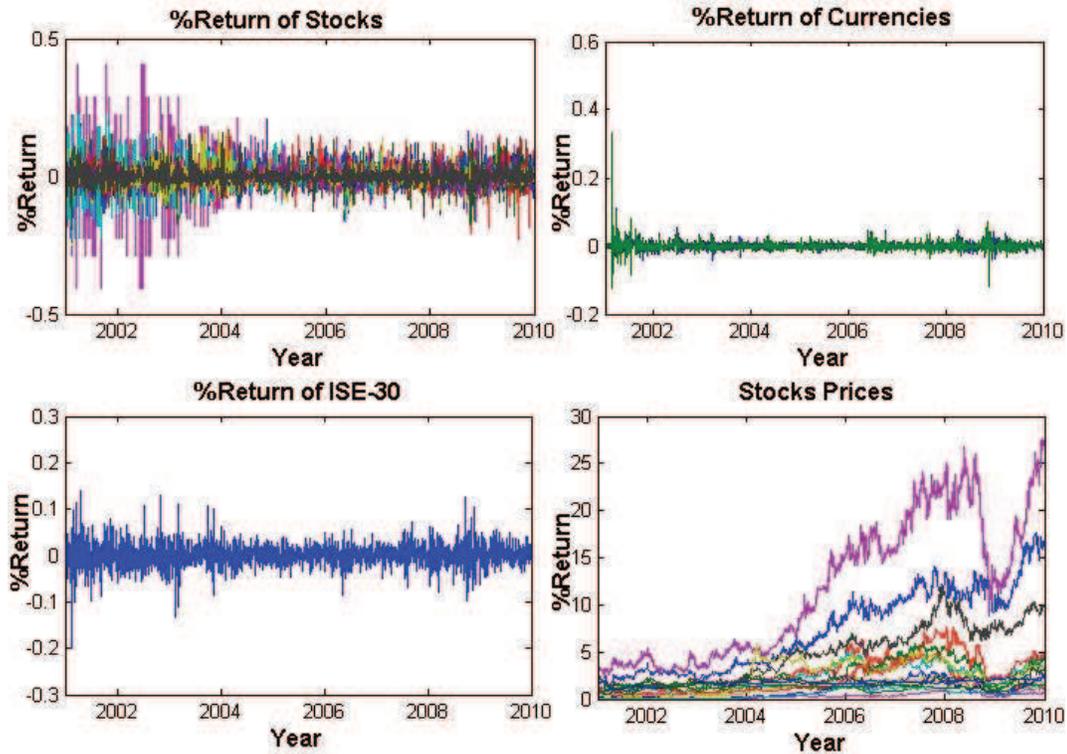}
\caption{Returns of Stocks, Currencies, ISE-30 as groups and Stocks Prices }
\end{center}
\end{figure}
From this figure, one can see that ISE 30 average returns and standard deviations are quite volatile between
2001 and 2003, then move almost horizontally until the second half of 2008 where the global crisis shows its effects.

\subsection{Observing Data Through Sliding Windows}

As discussed in the previous section, it is clear that the time
period 2001-2009 have regions of completely different character
and data should definitely be split into smaller portions.

We should remind that our emphasis is in obtaining a measure of the lengths of relevant
``past" and ``future" intervals with respect to portfolio optimization.  The key tool will be to
observe the past over a ``sliding window".  A sliding window is the observation of the data for past
$n$ days applied each day starting from the  $n+1$ day. If the past data is taken ``as is", i.e, with
a weight of $1$ we say that we observe the data over a ```rectangular window". Based on theoretical or
technical considerations, one can multiply the observed data by certain functions to get different
observation windows. As an example of technical requirements,  in the case of Fourier analysis,
one usually smooths out the discontinuities at the end points in order to get a better power spectrum.
In the case of financial data, one usually argues that past market conditions have less effect
on present and prefers to suppress past data usually by an exponential factor. In this work we
modify the observation window by multiplying with the
left  half of the  Gaussian Distribution Function (Normal Distribution) a
$$f(X)=\frac{1}{\sigma\sqrt{2\pi}} e^{-\frac{(x-\mu)^2}{2\sigma^2}}.$$

We have computed the mean and the variance of the market index over 10-day sliding windows as a
representation of almost instantaneous expected returns and standard deviations as displayed in Figure 3.

\begin{figure}[h]
\begin{center}
\includegraphics[width=375pt,keepaspectratio=true]{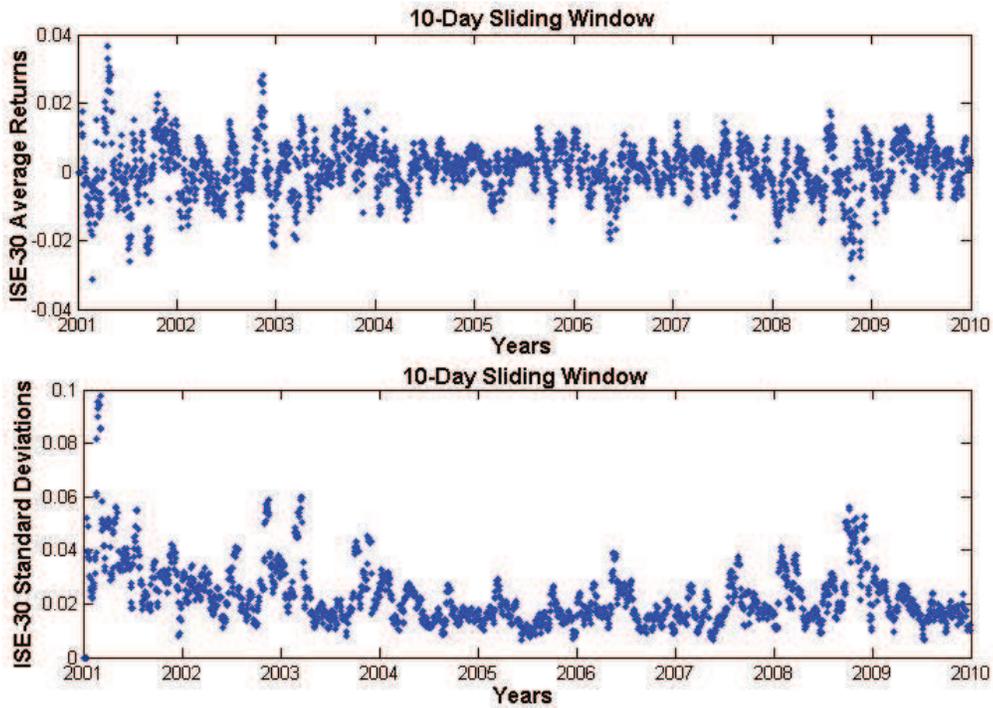}
\caption{10-day Sliding window return and standard deviation being reference}
\end{center}
\end{figure}

\section{Simulation Results}

In this section we first compare qualitatively the parameters as
seen from sliding windows of various lengths with the 10-day
sliding window is used as reference.  We then describe the trading
algorithm  and present the simulation results.

\subsection{Comparison of Sliding Windows}
 We now compare the means and
standard deviations of the ISE 30 index, computed using sliding
windows of various lengths with their instantaneous values
represented by 10-day sliding windows. We give below the
comparisons for 50 and 100 day observation periods that will be
used in trading.

\begin{figure}[!h]
\begin{center}
\includegraphics[width=425pt,keepaspectratio=true]{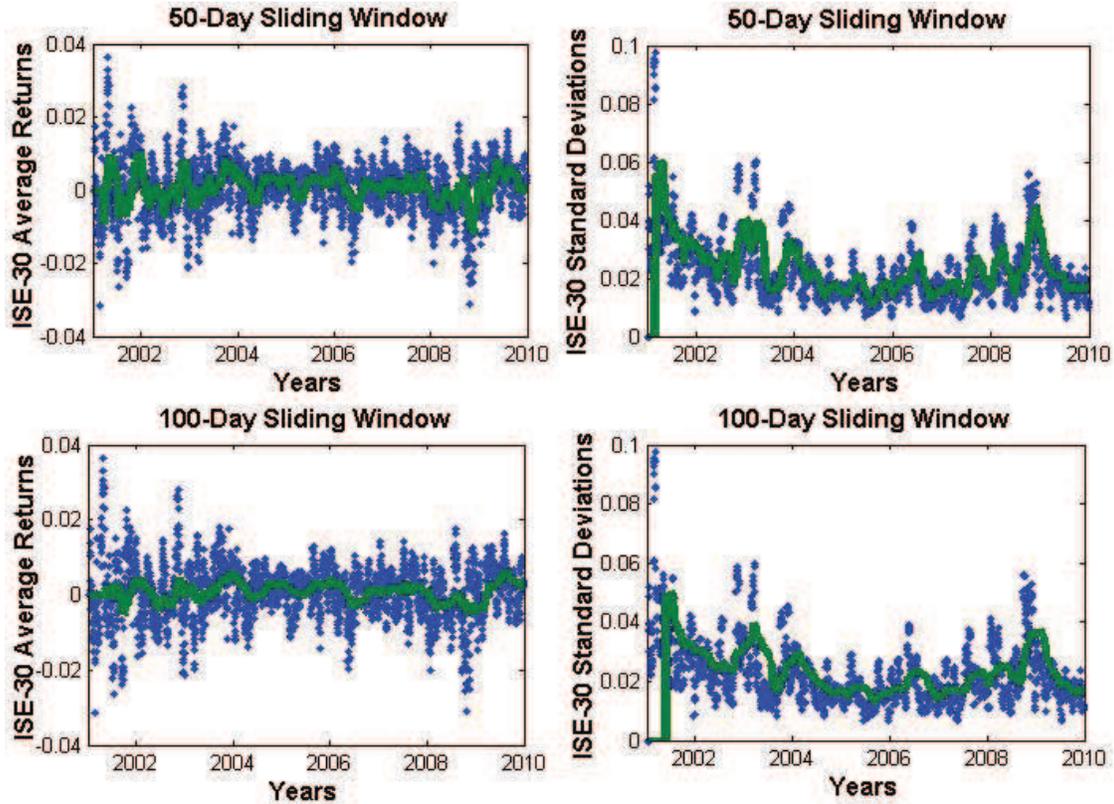}
\end{center}
\caption{50/100-day sliding window application the solid continuous (green) line shows
the 50/100-day period and the dots (blue points) show the 10-day sliding
window thought to be a representative of the instantaneous values}
\end{figure}

By comparing these figures, it can be seen that for data observed
through longer periods follow a smoother curve but they fail to
react to abrupt changes in the market conditions. It is also clear that
when the observation period gets longer, the statistical parameters follow smoother curve.
Furthermore the extreme values of average returns and standard deviations
turnout to be have lower values.

Sliding windows of  150-day, 200-day, 250-day and 300-day periods have been
studied but they are not included in this paper as they failed to give reasonable
results in trading simulations.

\subsection{The Investment Strategy}
The investment
strategy that we have chosen, aims to compare the relative lengths
of the observation and trading periods. It is based on the
assumptions below.
\begin{description}
  \item[1)] Observation period is past ``$p$" days; investment period is future ``$q$" days.
  At day  ``$i$, the investor  buys shares according to his/her observations for past ``$p$"
  days and  holds these shares for ``$q$" days. The  investor sells these  at the ``$(i+q)$th" day,
  regardless of the market conditions. Investor calculates the daily
return by dividing the difference between selling price and buying
price to buying price. And this result is divided by ``$q$"  to
obtain a daily average to be compared with other investment
periods' results.
  \item[2)] The average return vector and covariance matrix are computed by observing the past ``$p$" days.
  \item[3)] Target return is assumed to be a multiple of the ISE-30 index at day ``$i$.
  If return of the ISE 30 index is negative that day no purchase is made.
  \item[4)] The quadratic optimization program is run with these parameters. No short-sell is allowed.
  In there is no feasible solution, no purchase is made.
  \item[5)] There is no transaction cost. The sum of the weights of the instruments included in our portfolio is equal to 1.
\end{description}

\subsection{Simulation Results}

We have run the trading algorithm described above by choosing target returns that are $k$ times
the return of the ISE 30 index, for $k=2$, $k=3$ and $k=10$. We had also studied higher multiples
to see how the algorithm performs under extreme conditions but as $k=10$ is sufficiently representative
of the extreme conditions, these  are not included here.

We present in Figure 5 the graphs of the daily returns computed as described above, for $k=2$.
Depending on the parameters we set the green which shows the ISE 30 index performance
and the blue line shows the performance of our portfolio under the same conditions.
Here we see that the algorithm results in trading the most of the days and also extreme gains/losses occur
to our portfolio performance at 2003 and 2005, except further situations
our portfolio outperformed in the same way as the ISE 30 index as expected.

\begin{figure}[!h]
\begin{center}
\includegraphics[width=300pt,keepaspectratio=true]{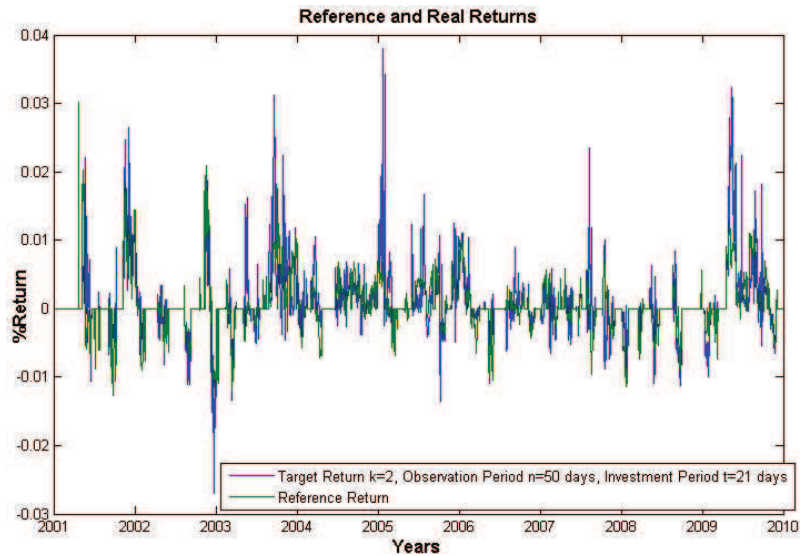}
\end{center}
\caption{Investment Strategy of the Program when Target Return k=2, Observations Period n=50days,   Investment Period t=21days  }
\end{figure}

Figure 6 displays the trading results for $k=10$, which is an extreme target return
we show how investment strategy acts if we only change the
rate of target return. For $k=10$ same computations are made and one can see that the
number of days which buy/sell operations are made decreased. In addition the volatility
and risk of operations increased. For the situation $k=2$, return values approximately
range from $-0.028$ to $0.039$ and for the situation $k=10$, return values approximately range from $-0.11$ to $0.13$.
 \begin{figure}[!ht]
 \centering
   \includegraphics[width=300pt,keepaspectratio=true]{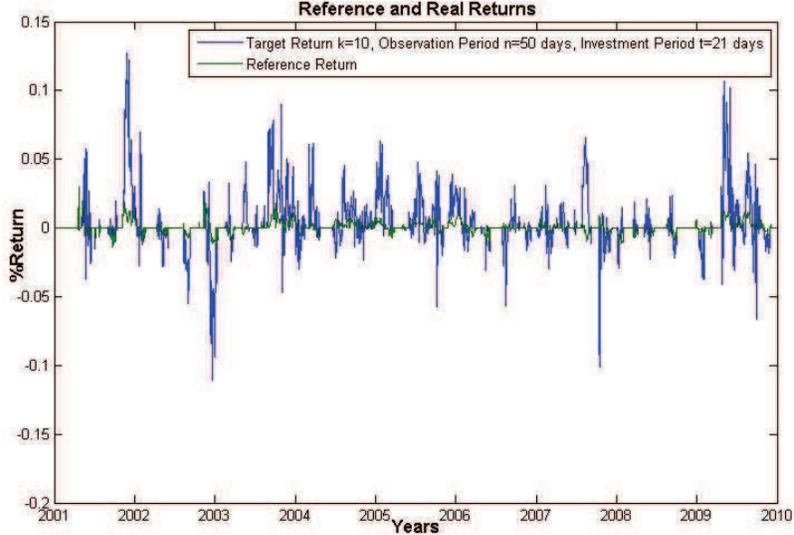}
 \caption{Investment Strategy of the Program when Target Return k=10, Observations Period n=50days, Investment Period t=21days }
\end{figure}

We have used observation and investment periods over a wide range but we just present the results that are meaningful in Table 1 below.
\begin{table} [!hb]
\caption{Average Reference and Real Return Rates Depending on Sliding Window Lengths}
\resizebox{16cm}{!} {
\centering 
\begin{tabular}{|c|c|c|c|c|c|c|c|c|c|c|}
\hline
\multicolumn{1}{|c|}{For k=2}& \multicolumn{10}{|c|}{Trading Period(Days)} \\
\hline
\multirow{2}{*}{Observation Period(Days)} &\multicolumn{2}{|c|}{21}& \multicolumn{2}{|c|}{42}&\multicolumn{2}{|c|}{63}&\multicolumn{2}{|c|}{84}&\multicolumn{2}{|c|}{105} \\ \cline{2-11}
 &\multicolumn{1}{|c|}{Reference}& \multicolumn{1}{|c|}{Real} & \multicolumn{1}{|c|}{Reference}& \multicolumn{1}{|c|}{Real} &\multicolumn{1}{|c|}{Reference}& \multicolumn{1}{|c|}{Real} &\multicolumn{1}{|c|}{Reference}& \multicolumn{1}{|c|}{Real} &\multicolumn{1}{|c|}{Reference}& \multicolumn{1}{|c|}{Real}  \\ \cline{1-11}
 {50}&\multicolumn{1}{|c|}{6.11}& \multicolumn{1}{|c|}{6.64} & \multicolumn{1}{|c|}{5.67}& \multicolumn{1}{|c|}{8.27} &\multicolumn{1}{|c|}{}& \multicolumn{1}{|c|}{} &\multicolumn{1}{|c|}{}& \multicolumn{1}{|c|}{} &\multicolumn{1}{|c|}{}& \multicolumn{1}{|c|}{}  \\
 {100}&\multicolumn{1}{|c|}{5.63}& \multicolumn{1}{|c|}{4.35} & \multicolumn{1}{|c|}{6.11}& \multicolumn{1}{|c|}{7.58} &\multicolumn{1}{|c|}{5.84}& \multicolumn{1}{|c|}{6.83} &\multicolumn{1}{|c|}{6.14}& \multicolumn{1}{|c|}{7.42} &\multicolumn{1}{|c|}{}& \multicolumn{1}{|c|}{}  \\
 {150}&\multicolumn{1}{|c|}{5.71}& \multicolumn{1}{|c|}{3.12} & \multicolumn{1}{|c|}{6.42}& \multicolumn{1}{|c|}{7.21} &\multicolumn{1}{|c|}{6.36}& \multicolumn{1}{|c|}{7.23} &\multicolumn{1}{|c|}{6.21}& \multicolumn{1}{|c|}{6.97} &\multicolumn{1}{|c|}{5.38}& \multicolumn{1}{|c|}{7.36}  \\
\hline\hline\hline

\hline
\multicolumn{1}{|c|}{For k=3}& \multicolumn{10}{|c|}{Trading Period(Days)} \\
\hline

\multirow{2}{*}{Observation Period(Days)} &\multicolumn{2}{|c|}{21}& \multicolumn{2}{|c|}{42}&\multicolumn{2}{|c|}{63}&\multicolumn{2}{|c|}{84}&\multicolumn{2}{|c|}{105} \\ \cline{2-11}
 &\multicolumn{1}{|c|}{Reference}& \multicolumn{1}{|c|}{Real} & \multicolumn{1}{|c|}{Reference}& \multicolumn{1}{|c|}{Real} &\multicolumn{1}{|c|}{Reference}& \multicolumn{1}{|c|}{Real} &\multicolumn{1}{|c|}{Reference}& \multicolumn{1}{|c|}{Real} &\multicolumn{1}{|c|}{Reference}& \multicolumn{1}{|c|}{Real}  \\ \cline{1-11}
 {50}&\multicolumn{1}{|c|}{6.11}& \multicolumn{1}{|c|}{4.68} & \multicolumn{1}{|c|}{5.67}& \multicolumn{1}{|c|}{6.33} &\multicolumn{1}{|c|}{}& \multicolumn{1}{|c|}{} &\multicolumn{1}{|c|}{}& \multicolumn{1}{|c|}{} &\multicolumn{1}{|c|}{}& \multicolumn{1}{|c|}{}  \\
 {100}&\multicolumn{1}{|c|}{5.63}& \multicolumn{1}{|c|}{2.27} & \multicolumn{1}{|c|}{6.11}& \multicolumn{1}{|c|}{4.9} &\multicolumn{1}{|c|}{5.84}& \multicolumn{1}{|c|}{5.85} &\multicolumn{1}{|c|}{6.14}& \multicolumn{1}{|c|}{5.56} &\multicolumn{1}{|c|}{}& \multicolumn{1}{|c|}{}  \\
 \hline\hline\hline

 \hline
\multicolumn{1}{|c|}{For k=10}& \multicolumn{10}{|c|}{Trading Period(Days)} \\
\hline

\multirow{2}{*}{Observation Period(Days)} &\multicolumn{2}{|c|}{21}& \multicolumn{2}{|c|}{42}&\multicolumn{2}{|c|}{63}&\multicolumn{2}{|c|}{84}&\multicolumn{2}{|c|}{105} \\ \cline{2-11}
 &\multicolumn{1}{|c|}{Reference}& \multicolumn{1}{|c|}{Real} & \multicolumn{1}{|c|}{Reference}& \multicolumn{1}{|c|}{Real} &\multicolumn{1}{|c|}{Reference}& \multicolumn{1}{|c|}{Real} &\multicolumn{1}{|c|}{Reference}& \multicolumn{1}{|c|}{Real} &\multicolumn{1}{|c|}{Reference}& \multicolumn{1}{|c|}{Real}  \\ \cline{1-11}
 {50}&\multicolumn{1}{|c|}{6.11}& \multicolumn{1}{|c|}{1.38} & \multicolumn{1}{|c|}{5.67}& \multicolumn{1}{|c|}{7.39} &\multicolumn{1}{|c|}{}& \multicolumn{1}{|c|}{} &\multicolumn{1}{|c|}{}& \multicolumn{1}{|c|}{} &\multicolumn{1}{|c|}{}& \multicolumn{1}{|c|}{}  \\
 {100}&\multicolumn{1}{|c|}{5.63}& \multicolumn{1}{|c|}{6.05} & \multicolumn{1}{|c|}{6.11}& \multicolumn{1}{|c|}{3.25} &\multicolumn{1}{|c|}{5.84}& \multicolumn{1}{|c|}{2.52} &\multicolumn{1}{|c|}{6.14}& \multicolumn{1}{|c|}{1.83} &\multicolumn{1}{|c|}{}& \multicolumn{1}{|c|}{}  \\
 \hline

\end{tabular}
}
\end{table}

The return rates given in this table are obtained by observing past $n=50, 100$ or $150$ days from a sliding window; the investment is then made to assets for investment periods of $t=21, 42, 63, 84$ or $105$ days and this operation is repeated for all applicable days between $2001$-$2009$.

One can see that the best result corresponds to the case where the investor aims $k=2$ times of reference return,
the observation period is 50-days and trading period is 42-days.
Even if exactly k-times of the targeted returns cannot be achieved, the real returns are very close to reference returns
or sometimes higher at lower targets. For higher returns the real
returns usually decrease in all cases.
\section{Conclusion}

In this paper we made a simulation for the performance of a portfolio under various
choices of target returns and investment periods, with the emphasis on the lengths of
 observation periods for the determination of statistical parameters.  Within the framework of our
 trading strategy,   we have tested extreme cases by aiming
target returns up to  $100$ times the market return and
we have seen that target returns should be a moderate multiple ($2$-$3$ times) of the market return in order to achieve
the aimed returns.

For each of these cases, we increased the observation periods from  $21$ days to $105$ days, by $21$-day steps,
and we increased the investment periods
$10$ days to $300$ days, by $10$-day steps, but we presented here only the ones that
 lead to most reasonable results.

We finally note that our investment strategy doesn't allow short selling and this has to be ensured in the simulations.

\end{document}